\documentstyle[12pt]{article}
\title{New topologies in the fundamental diagram of a cellular automata traffic flow model} 
  
\author{H. Ez-Zahraouy$^*$, K. Jetto, A. Benyoussef\\\
\\
 Laboratoire de Magn\'{e}tisme et de la Physique
 des Hautes Energies
\\
Universit\'{e} Mohammed V, Facult\'{e} des Sciences, Avenue Ibn Batouta,  B.P. 1014
\\Rabat, Morocco
}
\date{ }

\begin{document}
\maketitle 
\abstract{  
We propose a traffic flow model in which the vehicles are filed from their maximal velocities, the fast cars run with $Vmax{1}$, whereas the slow ones run with $Vmax{2}$. Using new overtaking rules which deals with deterministic NaSch model, it is found that the fundamental diagram exhibits three new topologies, depending on the fractions $f_{fast}$ and $f_{slow}$ of fast and slow vehicles respectively, in which the current profile displays two branches with negative slopes and two branches with positive ones. Moreover, in the second branch of the fundamental diagram, the model exhibits an absorbing phase transitions in which the behaviour of the order parameter $f_{d}$ and the current J is described by the power laws. In this case, it'is found that the system present a universel scaling law. On the other hand, a simple change in the rule of overtaking induce the metastability which depends on the state of the chain instead of external parameters \cite{4,5,6}. Furthermore, in the case of random fractions of vehicles, the fundamental diagrams are similar to the experiments results \cite{7,8}}\\.

Pacs numbers: 89.75.Fb , 05.65.+b, 05.70.Ln 
\\$ ^*$corresponding author: ezahamid@fsr.ac.ma  
\newpage
\section{\protect\bigskip Introduction}
The main raison for studying vehicular traffic  is its  no stop increasing  problems. Indeed, a new study suggests that traffic fumes is killing tens of thousands of people across Europe, and the costs of treating the illness associated with traffic pollution exceeds the costs arising from traffic accidents\cite{1}. On the other hand, the congestion traffic has a harmful impact on the economy since drivers and goods waste more time and fuel standing in a traffic jams( The Texas transportation institute found that the national cost of traffic congestion is 62.3 billion \$  annually)\cite{2}. Moreover, the construction of new highway is not affordable and publicly acceptable due to the expensive costs and for want of lands especially in urban areas. Many solutions and strategies were proposed in order to use the available infrastructures more efficiently, but with little success. For such raisons scientists including physicists, mathematicians, chemists and engineers are still working in this field in order to give a general solution, which can reduce the damage caused by roadway traffic.\\ 
Obviously , traffic flow depends strongly on the density C of cars on the road, but how vary the flux J and the average velocity V with vehicles density C? Functional relations between J, V, and C have been measured since 1935 by Greenshields (see\cite{9}). However, it is difficult to obtain very reliable and reproducible detailed empirical data on real traffic for several reasons namely: It is not possible to perform such laboratory experiments on vehicular traffic, thus empirical data are collected through passive observations rather than active experiments. Secondly the interpretation and evaluation of the collected data is not a simple task because traffic states depend on several influences like: weather conditions, presence of incidents, the measurement conditions( location where the traffic variable were measured, speed limits, ...), and other irregularities\cite{9,10} . As a consequence,  the empirical determination of the dynamical properties of traffic flow is not evident.  In order to use the empirical results for a theoretical analysis, many authors have used the mean values of the flow at given density which lead to a collection of possible forms of averaged fundamental diagrams(J,C) \cite{3,7} consistent with empirical data. The availability of these data has stimulated a great number of experimental studies and their comparison with theoretical models. The physical theory of traffic is based on the concept and techniques of statistical mechanics\cite{11,12,13}, the complex behaviour of traffic flow on a freeway has been the major topic of current interests \cite{14,15}. Indeed, many theoretical research using both analytical and numerical simulations in one and two dimension, has been done in order to make clear the characteristics of this complex behaviour \cite{16,17}. In this context there are  two different ways for modelling traffic: The macroscopic models which are based on fluid-dynamical description and the microscopic ones where attention is explicitly focused on individual vehicles which are represented by particles. The interaction is determined by the way the vehicles influence each others movement. In other words in the microscopic theories traffic flow is considered as a system of interacting particles driven far from equilibrium\cite{9}. Thus, it offers the possibility to study various fundamental aspects of the dynamics of truly non equilibrium systems which are of current interest in statistical physics\cite{9,18,19}. Within the conceptual framework of the microscopic approach, the particle hopping models describe traffic in terms of stochastic dynamics of individual vehicles which are usually formulated using the language of cellular automata (CA)\cite{20}. In general, CA are an idealization of physical systems in which both space and time are assumed to be discrete and each of the interacting cell can have only a finite number of discrete states, thus in CA models of vehicular traffic first proposed by Nagel and Schreckenberg\cite{21}and subsequently studied by other authors using a variety of techniques\cite{22,23},  the position, speed, acceleration as well as time are treated as discrete variables and the lane is represented by a one-dimensional lattice, each site represents a cell which can be either empty or occupied by at most one vehicle at a given instant of time. Hence, these properties lead to a complex dynamic behaviour and a fast simulation of a great number of interacting vehicles.\\
However, the vehicles on the road run with different speed. Thus, for  modelling  traffic with  different kind of vehicles, for example cars and trucks, we consider  $f_{fast}$ the fraction of fast vehicles(type1), i.e, vehicles with a higher maximal velocity $V_{max1}$, while the remaining fraction $1-f_{fast}$ is assigned to slow vehicles(type2), i.e, vehicles with low maximal velocity $V_{max2}$. Hence, for a fixed density C, we denote by $C_{fast}=C\:f_{fast}$ and $C_{slow}=C\:f_{slow}$ the density of type1 and type2 respectively. Such situation leads to the formation of coherent moving blocks of vehicles each of which is led by a vehicle with low     $V_{max}$, i.e., the phenomena of platoon formation \cite{24}. The strong effect of slow vehicles is also shown in the multi-lane systems \cite{25,26}, indeed, the simulation results revealed that already for small densities, the fast vehicles move with the average free flow velocity of the slow cars even if only a small fraction of slow vehicles have been considered. This effect is quite robust for different choices of C.A model as well as for different lane changing rules. In this paper, we propose a CA traffic flow model which we hope can contribute to a better understanding of the traffic flow. Our analysis is inspired by the situation in real traffic hence, the proposed model is based on the following facts: a mixture of fast and slow vehicles taking into account the disorder in their  maximal velocities, more real traffic overtaking rules which deals with deterministic NaSch model in a single lane\cite{21,23}, metastability, and random fractions of different kind of vehicles. For didactical reasons the paper is organized as follow; section 2 is devoted to explain the model. In section 3 we will present the main results obtained by simulation with a critical discussions including an analytic treatment, scaling analysis and the metastability. In the last section we will present a general conclusion. 
\section{Model}
It is well known that Nagel and Schreckenberg  are among the early pioneers who have addressed the problems of traffic flow; their CA model\cite{21} has stimulated a great numbers of publications(for review see\cite{9,10}). Indeed, the NaSch model is based on four steps which are necessary to reproduce the basic features of real traffic; however more additional rules should be added in order to capture more complex situations. In this context we have formulated our model. Indeed, the case of disorder on the maximal velocities  leads to platoon phenomena as was discussed in the introduction, this means that the fast vehicles(type1 running with $V_{max1}$) are trapped in a coherent moving blocks which are led by a slow vehicle(type2 running with $V_{max2}$). Thus, in order to overcome this situation fast vehicles must overtake the slow ones, which will lead to a competition between the formation of platoons and their dissociation by overtaking. However, the choice of the overtaking rules is not a simple task due to the strong effect of slow vehicles. The situation of overtaking is considered only when the following rules(which deal with deterministic NaSch model) are satisfied:\\\
\textbf{C1}: type1 follows type2\\\\
\textbf{C2}: $Gap1\leq Vmax{2}$\\\
Gap1 is the distance between the two vehicles type1 and type2. This choice ensures that fast vehicles driving in a slow platoon try to overtake if it is possible.\\\\
\textbf{C3}: $ Gap2\geq gap_C$\\\
Gap2 is the number of empty sites in front of type2, i.e., the gap between type2 and the next vehicle in the chain, while $gap_C=min(V(type2)+1,Vmax{2})+1$ is the minimal safety distance required for overtaking. Indeed, the vehicles in the chain move at the same time step(parallel dynamic), and according to the deterministic NaSch rules $V(type2)=min(V(type2)+1,Vmax{2},Gap2)$, hence, for large Gape2 the velocities of type2 in the next time step will be $V(type2)=min(V(type2)+1,Vmax{2})$, as a consequence \textbf{C3} ensures that at the situation of overtaking, type2 moves without any hindrance which means that type1 estimates the velocity of type2 before overtaking.\\\\
\textbf{C4}:  if $ p_{s} > rand $, then type1 overtakes with $V(type1)=min(NGap, Vmax{1})$.\\\
$p_{s}$ is the overtaking probability and Rand is a random number. On the other hand, to avoid collision with vehicles in front of type2: NGap=Gap1+1+Gap2 is the new gap required for overtaking which is obtained according to \textbf{C2} and \textbf{C3}.\
Thus when the above criteria are satisfied type1 overtakes with its maximal velocity if enough empty space is allowed.\\\\
\textbf{C5}:  $Vmax{1}\geq 2(Vmax{2}+1)$\\\
This inequality guides the choice of $Vmax{1}$ and $Vmax{2}$; it is obtained according to the following criteria \textbf{C2}, \textbf{C3} and \textbf{C4}.\\\\
In general, the update in this model is divided into two sub steps:\\ 
in the first sub step (\textbf{I}), type1 which satisfies the four criteria can overtake in parallel; However, in the second substep (\textbf{II}), the remaining vehicles (i.e, types1 which didn't satisfy the four criteria and types2) move forward in parallel according to the NaSch rules, without taking into account the new positions of the type1 moved in the first sub step.

\section{results and discussion}
\subsection{Simulation results }
In our computational studies we have considered a chain with $N=1000$ sites and periodic boundaries. In this case simulation begins with particles randomly distributed around the chain according to their densities $C_{fast}=Cf_{fast}$ and $C_{slow}=Cf_{slow}$ where C is the global density. The systems run for more than 40 000 MCS to ensure that steady state is reached, at this moment data including the current and velocity are collected. In order to eliminate the fluctuations about 100 initial configurations were randomly chosen.  

For $p_{s}=1.0$ i.e., deterministic case and depending on the values of $f_{fast}$, the fundamental diagram (J,C) exhibits three new topologies in which four different branches occur.  In fact, at free flow we have two branches: one with positive slope and the other with negative one, which meet at local maximal current. The same behaviour occurs at congested traffic flow. The maximal current at free flow can be smaller, equal or greater than the one located at congested traffic flow. Indeed, for $Vmax{1}=5$ and $Vmax{2}=1$ in Fig.1-a the fraction of the fast vehicles is smaller than the fraction of the  slower ones $f_{fast}\:<<\:f_{slow}$, then the maximum of the free flow branch is smaller than the one of the congested branch. For $f_{fast}\:>>\:f_{slow}$, the maximum of the free flow branch becomes greater than the congested branch (Fig.1-c), while for $f_{fast}\: \approx \:f_{slow}$ the two maximums are equal (Fig.1-b). Such topologies in the fundamental diagram was obtained in empirical work\cite{3}. Moreover, in Figs.1, we can distinguish several different states of traffic depending on the value of the density C of vehicles in the road:

For $C\leq C_{max1}$ (the first branch), the current increases with C to reach at $C_{max1}$ its first maximal value $J_{max1}$ which depend strongly on the values of $f_{fast}$, indeed,$(C_{max1},J_{max1})=\{(0.14,0.288);(0.16,0.48);(0.19,0.68)\}$ for $f_{fast}=\{0.3,0.6,0.75\}$ respectively. In this region, the four criteria of overtaking are satisfied due to the lower values of C, i.e, an enough empty space between vehicles, thus all vehicles type1 overtake type2 which means that both types of vehicles are in their free flow regime as shown in Fig.2.

For $C_{max1}\leq C\leq C_{min}$ (the second branch); $C_{min}$ is the density at which the current show a nonzero minimum $J_{min}$. In this region the current value fall with C. Indeed, the situation of overtaking becomes more difficult as long as C increases, thus not all vehicles type1 can overtake, which leads to a competition between the formation of platoons and their dissociation by overtaking; as a consequence type1 are in their congested phase whereas type2 are still in their free flow regime as displayed in Fig.2.

For $C_{min}\leq C\leq C_{max2}$ (the third branch) type1 can not overtakes type2, such situation leads to the formation of coherent moving blocks of vehicles each of which is led by type2, i.e, the phenomena of platoon formation.
 
We recall that $C_{max2}$ is the density at which the current show a second maximal value $J_{max2}$. Thus for $C \geq C_{max2}$(the fourth branch) both vehicles are in the jamming phase.

In Fig.2 we have plotted the mean speed of each types of vehicles versus the density of cars C with the same fractions as in Fig.1-c. Hence Fig.2 proves that for $(C\leq C_{max1})$ (the first branch) each types of vehicles run with their average free flow velocity. For $C_{max1}\leq C\leq C_{min}$ (the second branch) the average speed of type1 fall with C while the average speed of type2 remain constant, which means that types2  are still in their free flow regime whereas types1 are in their congested phase. For $C_{min}\leq C\leq C_{max2}$ (the third branch) type1 and type2 run with the same speed .i.e., type1 drive in a slow platoon which is led by type2; While for $C \geq C_{max2}$ the jamming phase take place.

We conclude that for type1 the first maximum $(C_{max1},J_{max1})$ corresponds to the transition from free flow to congested phase. $(C_{min},J_{min})$ indicates the transition from congested phase to platoon phase while the second maximum $(C_{max2},J_{max2})$ corresponds to the transition from platoon phase to jamming phase of both types of vehicles.\\
On the other hand, the same analysis hold for the case when $p_{s} < 1.0$ except that $J_{max1}$ is lowered whereas $C_{max1}$ and $C_{min}$ are shifted to the left as displayed in Fig.3. Indeed, even if enough empty space exist not all type1 can over take due to \textbf{C4}.\\ 
Moreover, $J_{min}$ does not depend on $f_{fast}$ and $p_{s}$, in contrast to $J_{max1}$. This is due to the structure of the chain in this region, i.e., there is no sufficiently empty space for overtaking in comparison to the first region. Also $C \geq C_{min}$ imply the end of the competition, hence, the movement of type1 is dictated by type2 which leads to the well known fundamental diagram of one species of vehicles with $V_{max}=1$, for such raison $J_{max2}$ is also independent on $f_{fast}$ and $p_{s}$.\\
Furthermore we believe that our model can be extended to the case of n types of vehicles with n different maximal velocities. In this context we have computed the case of three types of vehicles, where $f_{fast1}$, $f_{fast2}$ and $f_{slow}$ are the fractions of type1 ($V_{max1}=10$), type2 ($V_{max2}=4$) and type3 ($V_{max3}=1$) respectively. The choice of $V_{max}$ follows the condition \textbf{C5} mentioned above. In this case type1 can overtakes type2 and type3 whereas type2 can overtakes only type3. Hence, the current displays three local maximums. At free flow the maximums are comparable because both type1 and type2 are fast. Indeed, Fig.4 shows that for low density the three vehicles are in their free flow regime, however, the mean speed of type1 falls quickly with C in comparison with the average speed of type2, which explains the appearance of the first and second maximum. As long as C increases type1 and type2 are not able to overtake, hence, they drive in a slow platoon which is led by type3. 
\subsection{Analytic treatment }
As was discussed in the previous section the disorder in the maximal velocities and the overtaking rules lead to a competition between the platoons formation and their dissociation. Thus we need a physical quantity to quantify this competition. \\
Lets consider  $N_{f}$ the number of vehicles type1 in the steady state, for which the rules  \textbf{C1},\textbf{C2} and \textbf{C3} are satisfied .i.e., fast vehicles which are  favourable to overtake. Unfortunately,  not all the $N_{f}$ vehicles can do because \textbf{C4} will not be fulfilled for all these cars. As a consequence,  among  $N_{f}$ only I vehicles succeed to overtake depending on the value of $p_{s}$. The probability of such event is given by: 
\begin{equation}
           \textit{P}=C^I_{N_f} p_{s}^I (1-p_{s})^{N_f-I}
\end{equation}
With $$\:\:C^I_{N_f}=\frac{\rm N_{f}!}{\rm (N_f-I)!\:I!}$$
Hence, the mean value of I is: 
\begin{equation}
                     <I>=\sum_{i=1}^{N_f}i \: C^i_{N_f} p_s^i (1-p_{s})^{N_f-i}
\end{equation}
However, 
\begin{equation}
                 i \: C^i_{N_f}=N_{f} C^{i-1}_{N_f-1} 
\end{equation}
Thus,
\begin{equation}
       <I>=N_{f} p_{s}\:\sum_{i=1}^{N_f} C^{i-1}_{N_f-1} p_{s}^{i-1} (1-p_{s})^{(N_f-1)-(i-1)}
\end{equation}
for $i\rightarrow i-1$, the sum becomes a binomial law:
\begin{equation}
                \sum_{i=0}^{N_f-1} C^i_{N_f-1} p_{s}^i (1-p_{s})^{(N_f-1)-i}=1
\end{equation}
As a consequence, 
\begin{equation}
                <I>=N_{f}\: p_{s}
\end{equation}
On the other hand, we denote by $f_{d}$ the fraction of vehicles type1 which succeed to overtake:
\begin{equation}            
                f_{d}=\frac{\rm N_{f}}{\rm N_p}\: p_{s}
\end{equation}

Where, $N_{p}$ is the number of vehicles type1 which are trapped behind a slow vehicles .i.e., the number of fast vehicles for which at least \textbf{C1} is satisfied. Thereafter, $f_{d}$ is the physical quantity which quantify the competition between overtaking and platoons formation, indeed:

For $C\leq C_{max1}$  (the first branch)where there is enough empty space for overtaking, $f_{d}$ vehicles type1 run with $V_{max1}$ whereas the remaining fast vehicles (1-$f_{d}$) moves with the same velocity $V_{max2}$ as type2; Thus the current J can be expressed as:
\begin{equation}
               J(C)=C_{fast}[f_{d}\:V_{max1}+(1-f_{d})\:V_{max2}]+C_{slow}\:V_{max2}
\end{equation}

We recall that $C_{fast}=C\:f_{fast}$ and $C_{slow}=C\:f_{slow}$ are the density of type1 and type2 respectively, and $f_{fast}=1-f_{slow}$, hence, one can easily get:
\begin{equation}
    \:\: J(C)=C\:[f_{fast}\:f_{d}(V_{max1}-V_{max2})+V_{max2}]
\end{equation}
Furthermore, in this range of density, $N_{f}=N_{p}$ which leads to  
\begin{equation}
                                     f_{d}=p_{s}
\end{equation}
Hence, the current in the first branch is:
\begin{equation}
            \:\: J(C)=C\:[f_{fast}\:p_{s}(V_{max1}-V_{max2})+V_{max2}]
\end{equation}

For $C_{min}\leq C\leq C_{max2}$ (the third branch) where overtaking is note allowed, $N_{f}<< 1$ thus  
\begin{equation}
                        f_{d}=0
\end{equation}

which leads to current in the platoon phase:    
\begin{equation}
                         \:\: J(C)=C\:V_{max2} 
\end{equation}
For $C \geq C_{max2}$(the fourth branch), here the current is limited by the density of holes:
\begin{equation}
                       \:\: J(C)=1-C
\end{equation}

which is a characteristic of the jamming phase. However it is difficult to do analytical calculation for $N_{f} and\: N_{p}$ in the second branch $C_{max1}\leq C\leq C_{min}$, indeed the simulation results Fig.5 shows that $f_{d}$ is a singular function in this range of density even in the deterministic case $\:p_{s}=1.$, hence, the current in this branch keep the formula (9) where $f_{d}$ can be obtained by simulation. Furthermore, we remark that the expressions of the current above are obtained depending on the value of $f_{d}$ in each branch except the last one (jamming phase), thereafter, the fundamental diagram(J,C) Fig.1-c is given by:\\
\begin{equation}
                      \:\:J(C)=Min\{C[f_{d}\:f_{fast}(V_{max1}-V_{max2})+V_{max2}],1-C\}
\end{equation}
This general formula sum up the previous expressions, however, it'is not possible to obtain $(C_{max1},J_{max1})$ and $(C_{min},J_{min})$ by a simple derivative due to the fact that $f_{d}$ is a singular function in the second branch. Thus, in the following section we will present a new way to study this singular behaviour. 
\subsection{Power laws and scaling analysis}
It' is well known that in equilibrium systems, there is a certain lower dimension below which the system is always disordered. However, the situation is very different in nonequilibrium systems. This fact has initiated intense research regarding the phase transitions in one-dimensional driven diffusive systems far from equilibrium\cite{9}. Moreover, the concepts and techniques developed for equilibrium phase transitions can also be applied to nonequilibrium systems such as the order parameter concept, scaling, universality as well as renormalization group, therefore, numerous nonequilibrium critical phenomena were successfully investigated by such methods\cite{27}. Here we are interested in the so called Absorbing Phase Transitions(APT)\cite{27,28,29,30} which occur in nonequilibrium dynamical systems that are characterized by at least one absorbing state. Indeed, in our model the platoon phase(the third branch) is an absorbing state in which the fast vehicles become trapped forever, whereas the active phase correspond to free flow phase of type1(the first branch). The competition between the proliferation and the annihilation process which constitute the essential physics of the AFT, is described in this model by the competition between the formation of platoons(proliferation process) and their dissociation by overtaking(annihilation process). However, in the active phase($C\leq C_{max1}$) the annihilation (dissociation of platoons) prevail the proliferation(platoons formation) thus, $f_{d}$ the physical quantity which quantify this competition have a non zero value($f_{d}=p_{s}$). In the absorbing state($C_{min}\leq C\leq C_{max2}$) in which proliferation outweight annihilation, $f_{d}=0$. For $C_{max1}\leq C\leq C_{min}$ the system undergoes an  absorbing phase transition from the active phase to the absorbing phase where $f_{d}$ fall with a singular manner as long as C approach the critical point $C_{min}$. As a consequence, $f_{d}$ represent the order parameter of the system and the density C is the control parameter of the absorbing phase transition.\\
Moreover, it' is well known that critical systems are characterized by power laws which describe the singularities observed sufficiently close to the critical point; fortunately the singularity in this model appears in a specific region of the density i.e., when $C_{max1}\leq C\leq C_{min}$, furthermore, Fig.5 show that this region depend only on $f_{fast}$; Thus the singular part of  the order parameter is assumed to scale asymptotically as:
\begin{equation}
                 	\:\:f_{d}(C)\: \sim \:f_{fast}^\gamma g(C/f_{fast}^\delta)
\end{equation}
Where $g(x)=cte$ is a scaling function, which means that the singular part of $f_{d}$ is asymptotically a generalized homogeneous function\cite{31}. Indeed, the simulation results Fig.6 confirm this assumption:

For $f_{fast} \geq 0.6$
 $$\:\gamma=0 \:and\: \delta=0.773\:$$
However, for $f_{fast}< 0.6$ the singular part of $f_{d}$ is independent of $f_{fast}$ as shown in inset of Fig.6.i.e., both $\gamma=0 \:and\: \delta=0$.\\
On the other hand, the same analysis hold for the second branch of the fundamental diagram. Indeed, in section \textbf{3.1} we have shown that $(C_{max1},J_{max1})$ depend on both $f_{fast}$ and $p_{s}$, as a consequence the current in the second branch is also asymptotically a generalized homogeneous function, hence for $p_{s}=1$ it' is assumed to scale asymptotically as:
\begin{equation}
                            \:\:J(C)\: \sim \:f_{fast}^\alpha f(C/f_{fast}^\nu)
\end{equation}
with the scaling function $$f(x)=cte$$.                    
                          
This result is confirmed by simulation as displayed by Fig.7. However, the exponent $\alpha$ and  $\nu$ depend strongly on the value of $f_{fast}$:

For $f_{fast} \geq 0.6$,i.e., when the vehicles type2 are considered as an impurity $f_{fast}\:>>\:f_{slow}$:
                            $$\alpha=1.631\: \:and\:  \: \nu=0.773$$

For $0.4 \leq f_{fast} \leq 0.6$, here both type of vehicle are comparable $f_{fast} \approx f_{slow}$: 
                           $$\alpha=1.3\: \:and\:  \: \nu=0$$

For $f_{fast} \leq 0.4$, in this case the vehicles type1 are now considered as an impurity $f_{fast}\:<<\:f_{slow}$:
                            $$\alpha=0.532\: \:and\:  \:\nu=0$$
However, for a fixed fraction of type1 $f_{fast}$ and $p_{s}<1$ (the stochastic case) Fig.7, the current scale asymptotically as:
\begin{equation}
                (8)\:\:J(C)\: \sim \:p_{s}^{z} h(C/p_{s}^\eta)
\end{equation}
with the scaling function $$h(x)=cte$$.

Here we can distinguish two regions depending on the value of $p_{s}$, indeed:

For $0.8\leq p_{s}\leq 1$ $\:(p_{s}\longrightarrow 1)$ i.e., the system tend to the deterministic case:
                        $$z=0.9613\: \:and\:  \: \eta=1.0591$$

Whereas for $p_{s}\:<\:0.8$:
                        $$z=0.302\: \:and\:  \: \eta=0$$  
According to these results we can formulate some remarks:

Firstly, the critical exponents $\gamma, \delta, \alpha, \nu, z, \eta$ are not universal because they depend on the parameters of the systems. However, the model shows a universel scaling law $\:\delta=\nu\:$, which confirm that the behaviour of the system in the second branch of the fundamental diagram is imposed by the one of $f_{d}$ in this range of density. Secondly, $f_{fast}=\{0.6;0.4\}$ represent a crossover, indeed, the system change its critical behaviour depending on which types of vehicles is the majority, such result was mentioned in beginning of section \textbf{3.1}. Furthermore, the strong effect of $p_{s}$ appears only when $p_{s}\:<\:0.8$.     

\subsection{Metastability }
According to the previous sections, there is a density region $C_{max1}\leq C\leq C_{min}$(the second branch) in which we can have both the free flow and the congested traffic Fig.2. Such remark indicate that the system can exhibits metastability and the related hysteresis effects \cite{7}. This phenomena have been observed in coupled-map lattice models \cite{23}, and in some generalizations of the NaSch model which are based on slow to start rules\cite{4,5,6}. However, in the latter cases, the slow to start rule is formulated by a modifications of the braking steps of the original NaSch model, whereas in our model it is obtained by a simple change in the second sub step (\textbf{II}) of updating. Indeed, when \textbf{C3} is not satisfied, the corresponding vehicle type1 stop in this time step, it will wait until the next time steps to move forward according to NaSch rules. Hence our slow to start rule depends on \textbf{C3} ,i.e, on the configuration of the chain instead of external parameters as in \cite{4,5,6}. Starting from two different initial conditions, the mega jam (a large compact cluster of standing vehicles) and the homogeneous state (vehicles are distributed periodically with the same constant gap) we obtain the hysteresis observed in Fig.8.\\
\section{Conclusion}
The investigations made in the above sections lead to the following conclusion:\\
We have studied the case of mixture of fast and slow vehicles taking into account the disorder in their maximal velocities, the key ingredient of our proposed model is how the faster cars can overtake the slow ones. Indeed, using a new overtaking rules which deal with deterministic NaSch model in a single lane, we have found that the proposed model belong to the class of nonequilibrium dynamical systems which exhibits an Absorbing Phase Transitions(APT). This approach has better help to understand the main branches of the obtained fundamental diagram especially the second branch. Indeed, in this range of density the strong correlation between the movement of both types of vehicles depend on $p_{s}$ and $f_{fast}$, as a consequence the behaviour of the order parameter $f_{d}$ and the current J is described by the power laws(16-17-18). The critical exponents $\gamma, \delta, \alpha, \nu, z, \eta$ are not universal because they depend on the parameters of the systems. However, the model shows a universel scaling law $\:\delta=\nu\:$. Furthermore, we have extended the model to the case of three kinds of vehicles in which we have shown that the fundamental diagram exhibits three maximal currents; such result can be considered as an open question for experiments.\\
Secondly, we recall that our aim is to propose a CA traffic flow model which we hope can contribute to a better understanding of the traffic flow. Hence, according to the previous discussions in section \textbf{3} we can say that our proposed model can give a pertinent indications about empirical results. Indeed, Figs.1 show the existence of two branches with positive slops which can coexist at the same density interval as shown in Fig.8. Such results was obtained in empirical work\cite{3}. moreover, in real traffic the fractions of each types of vehicles is unpredictable. Thus, for a fixed density C, there is an infinity of couples $(f_{fast},f_{slow})$ which can give different values of the current. Thus, according to this observation  we have studied the case of three types of vehicles where their fractions $(f_{fast1},f_{fast2},f_{slow})$ are chosen randomly. In this situation we have combined the disorder in the maximal speed, the disorder in the fraction of cars and the condition under which we can have the metastability. Hence, the results displayed in Fig.9 look similar to the empirical data \cite{7,8}.

\section*{Figure captions}
Fig.1-a: The fundamental diagram (J,C) for two types of vehicles in deterministic case $(p_{s}=1)$ with $(Vmax{1}=5,f_{fast}=0.3)$ for type1, and $(Vmax{2}=1,f_{slow}=0.7)$ for type2.\\\
Fig.1-b: The fundamental diagram (J,C) for two types of vehicles in deterministic case $(p_{s}=1)$ with $(Vmax{1}=5,f_{fast}=0.6)$ for type1, and $(Vmax{2}=1,f_{slow}=0.4)$ for type2.\\\
Fig.1-c: The fundamental diagram (J,C) for two types of vehicles in deterministic case $(p_{s}=1)$ with $(Vmax{1}=5,f_{fast}=0.75)$ for type1, and $(Vmax{2}=1,f_{slow}=0.4)$ for type2. The inset show the agreement between the analytic formula (14)(circle dots), and numerical similation (up triangles).\\\
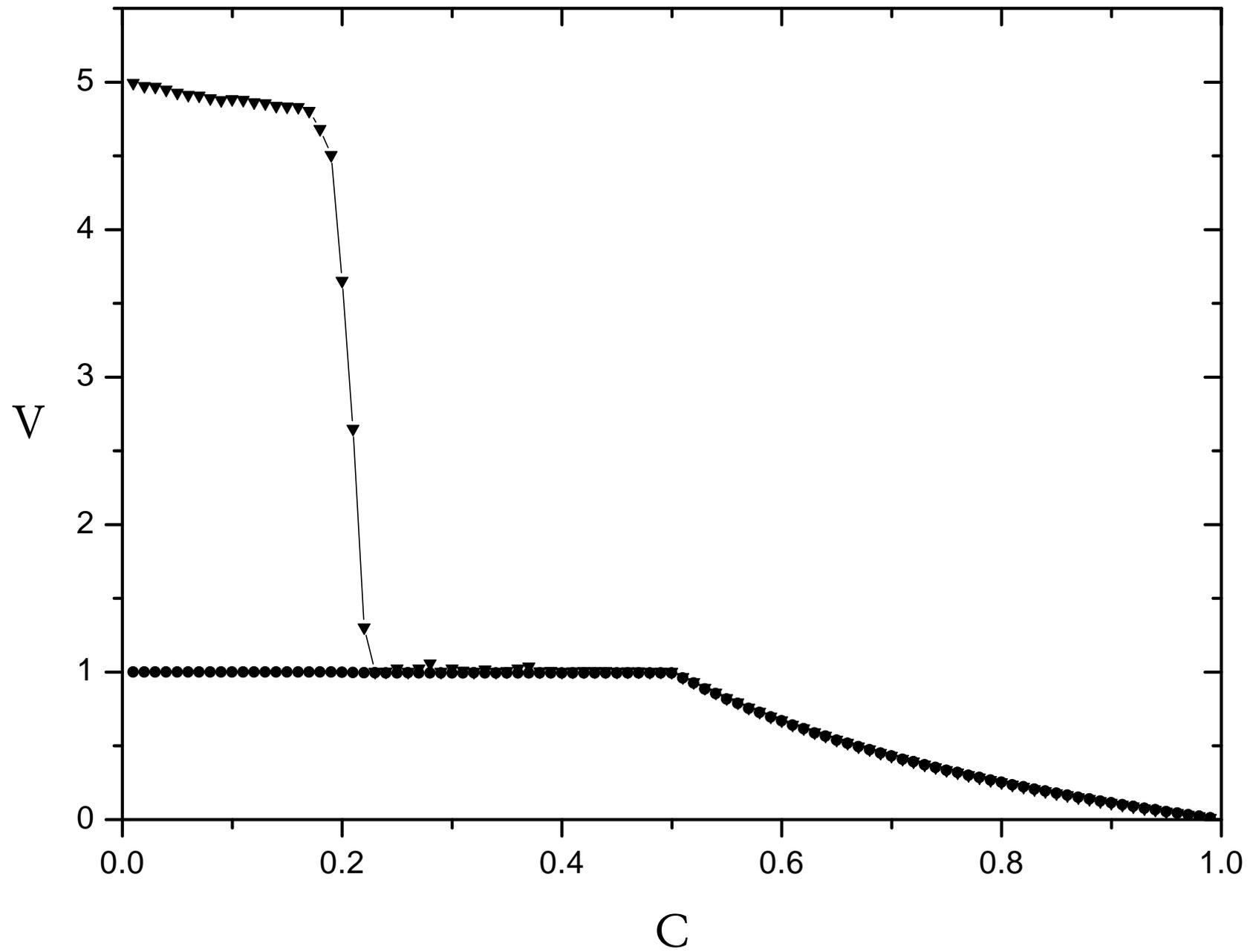
Fig.2: The mean speed V versus the density C for two types of vehicles, circle dots correspond to type2 with $(Vmax{2}=1,f_{slow}=0.25)$, whereas down triangles correspond to type1 $(Vmax{1}=5,f_{fast}=0.75)$. These curves were obtained in deterministic case $(p_{s}=1)$.\\\
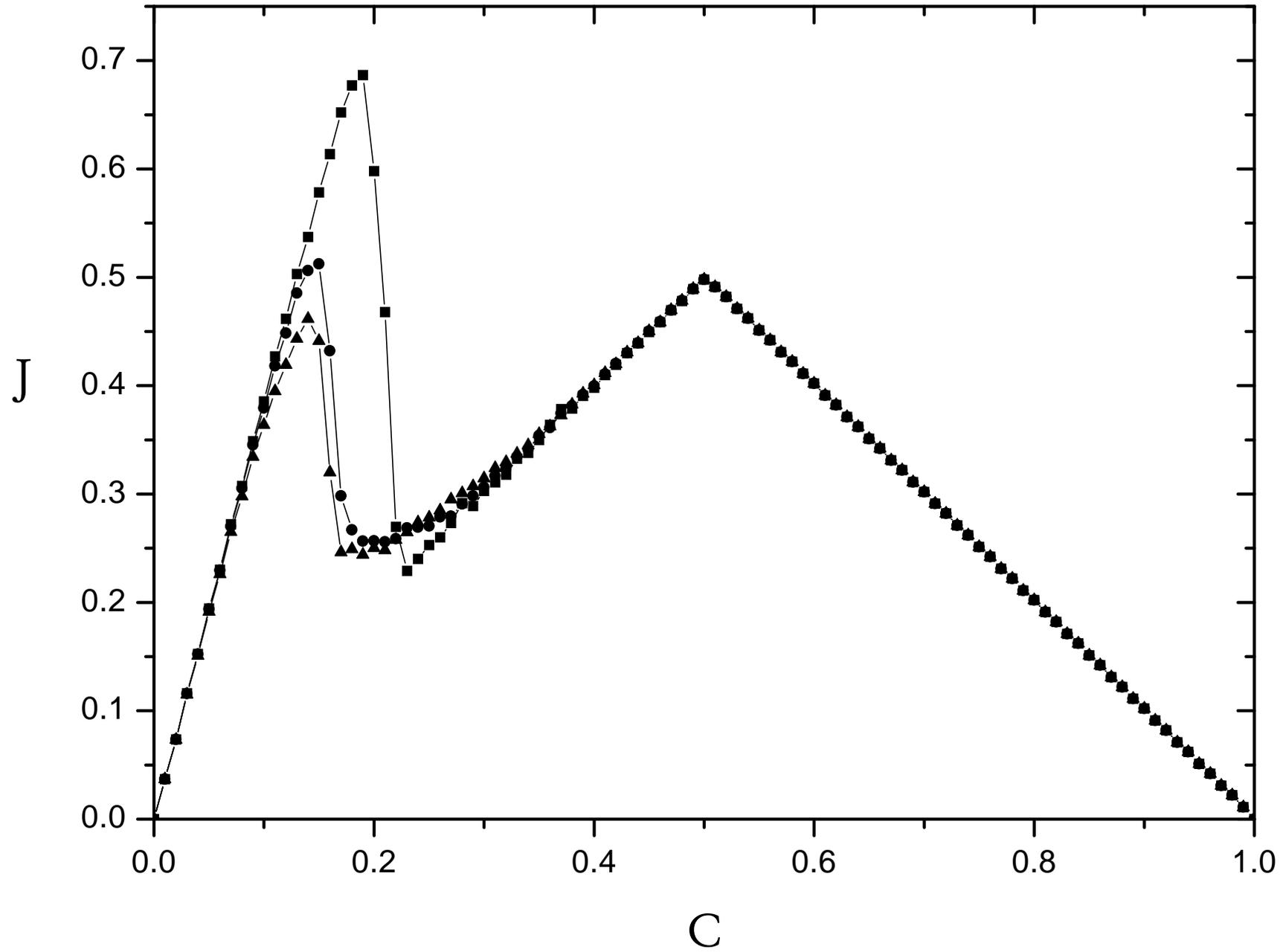
Fig.3: The fundamental diagram (J,C) for two types of vehicles with $(Vmax{1}=5,f_{fast}=0.75)$ for type1 and $(Vmax{2}=1,f_{slow}=0.25)$ for type2; circle dots correspond to deterministic case $p_{s}=1$, square dots $(p_{s}=0.7)$ and up triangles $(p_{s}=0.5)$ correspond to stochastic case.\\\ 
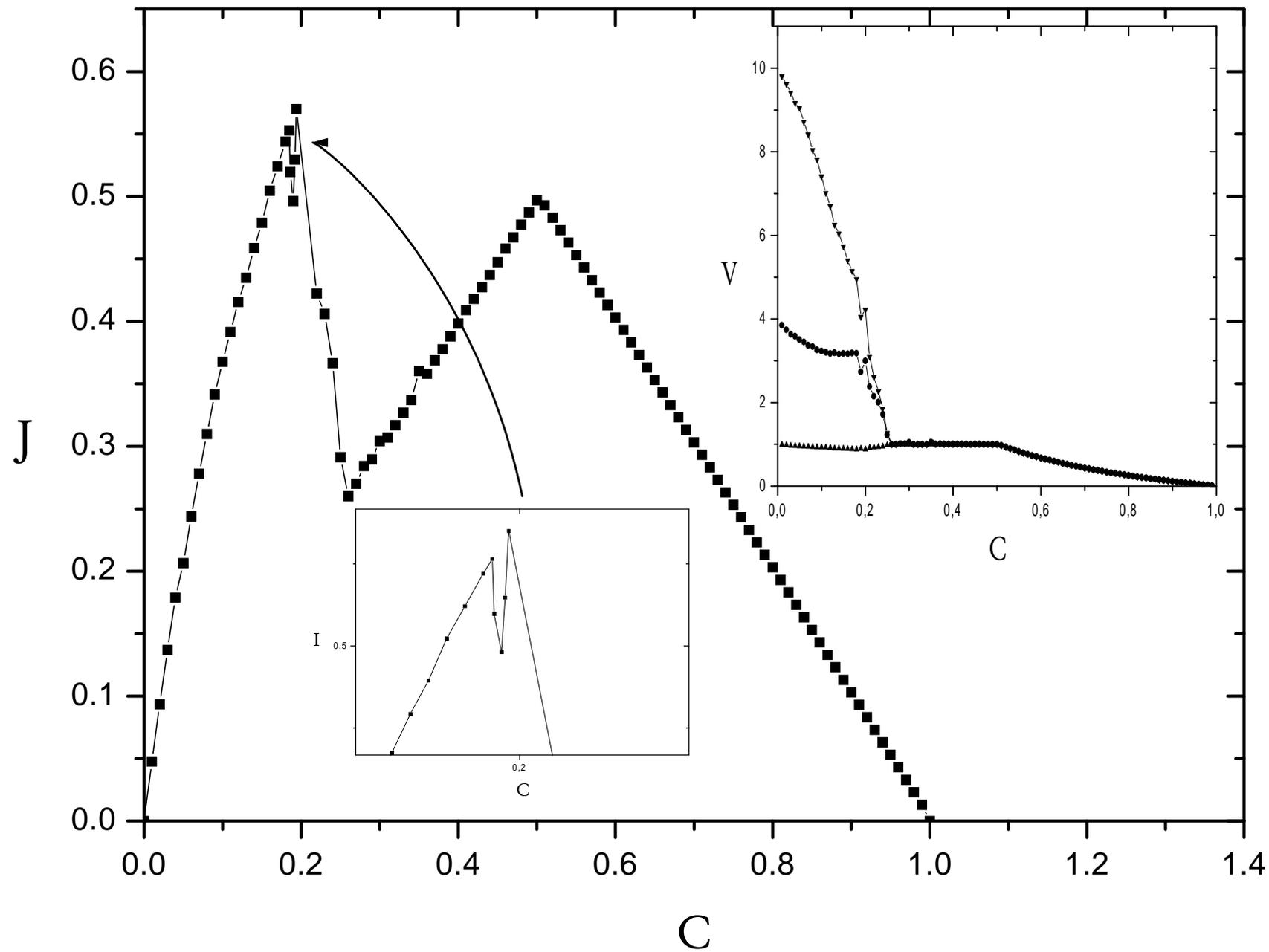
Fig.4: The fundamental diagram (J,C) and the mean speed V versus the density C (inset)for three types of vehicles in deterministic case $(p_{s}=1)$. With $(Vmax{1}=10,f_{fast1}=0.3)$ for type1(down triangles), $(Vmax{2}=4,f_{fast2}=0.4)$ for type2 (circle dots), and $(Vmax{3}=1,f_{slow}=0.3)$ for type3 (up triangles).\\\
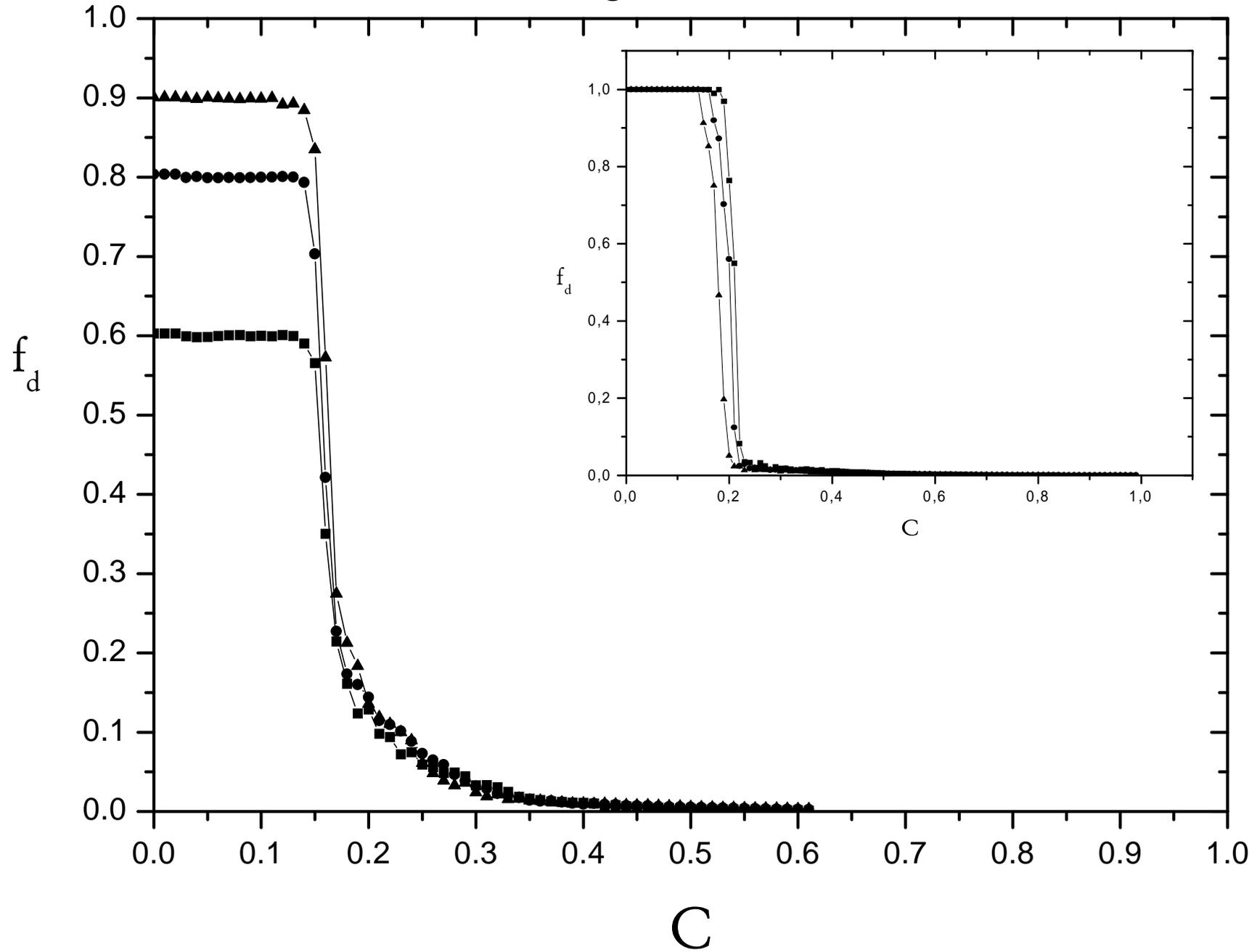
Fig.5: The fraction $f_{d}$ of fast vehicles which succeed to overtake versus the density C. In the main plot, the simulation was done  for two types of vehicles with $(Vmax{1}=5,f_{fast}=0.75)$ for type1, $(Vmax{2}=1,f_{slow}=0.25)$ for type2, and for three values of $p_{s}$: up triangles correspond to $p_{s}=0.9$, circle dots correspond to $p_{s}=0.8$, and $p_{s}=0.6$ for square dots. We remark that the analytic formulas(10-12) are verified.\\
However, in order to show the effect of $f_{fast}$ on the behaviour of $f_{d}$ (inset), the simulation was done in the deterministic case $p_{s}=1$, with $V(max{1}=5$ for type1, $Vmax{2}=1$ for type2, and for three values of $f_{fast}$: up triangles correspond to $f_{fast}=0.6$, circle dots correspond to $f_{fast}=0.7$, and $f_{fast}=0.75$ for square dots.\\\
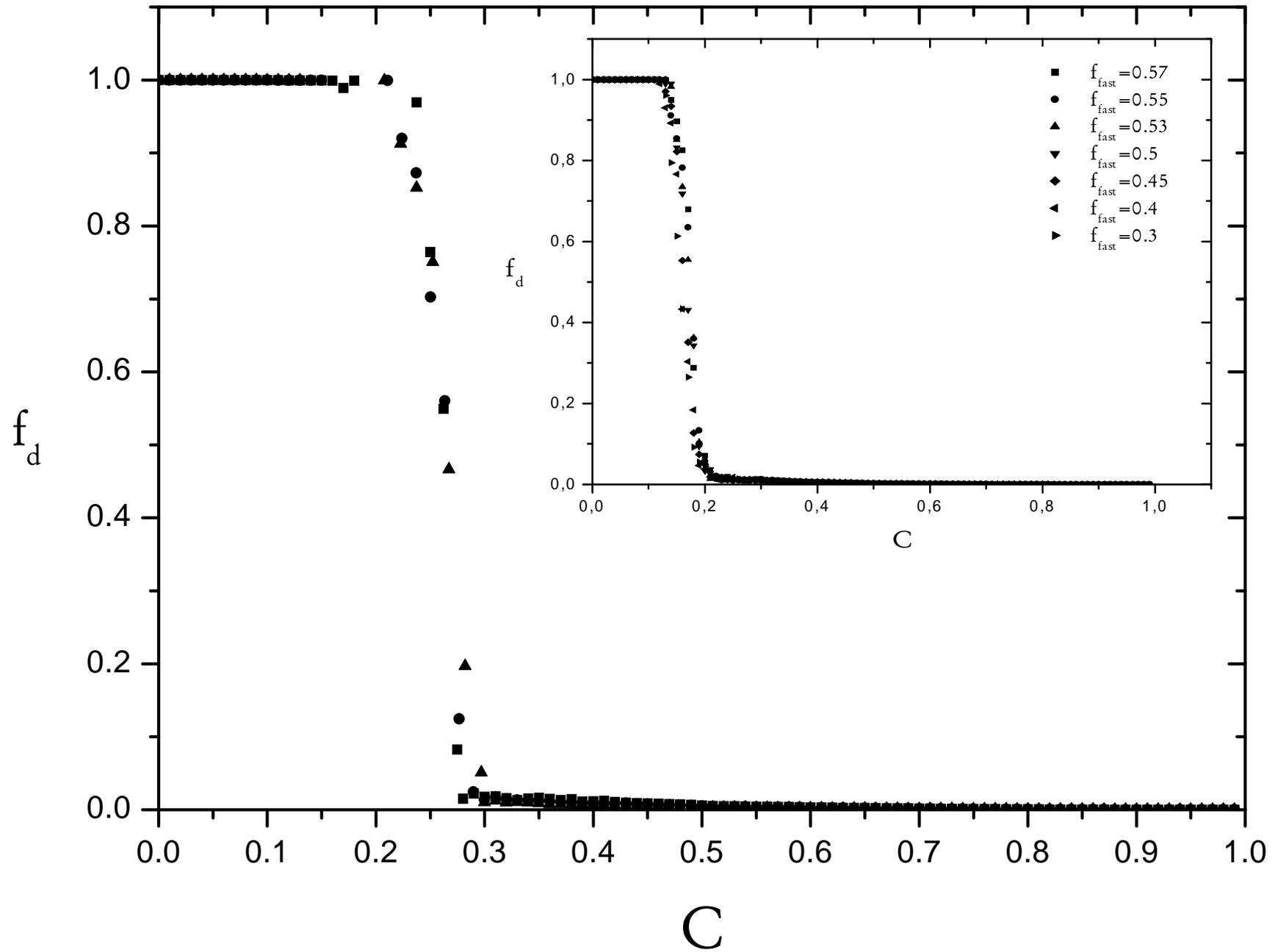
Fig.6: The scaling plot of the coexistence curve of the order parameter $f_{d}$ for three values of $f_{fast}$: square dots correspond to $f_{fast}=0.75$, circle dots correspond to $f_{fast}=0.7$, and up triangles correspond to $f_{fast}=0.6$. The case when $f_{fast}< 0.6$ is displayed in the inset. We note that the scaling is valid only in the singular part of the order parameter $f_{d}$ .i.e, for $C_{max1}\leq C\leq C_{min}$. The simulation was done for two types of vehicles in the deterministic case $p_{s}=1$ with $Vmax{1}=5$ for type1, and $Vmax{2}=1$ for type2.\\\
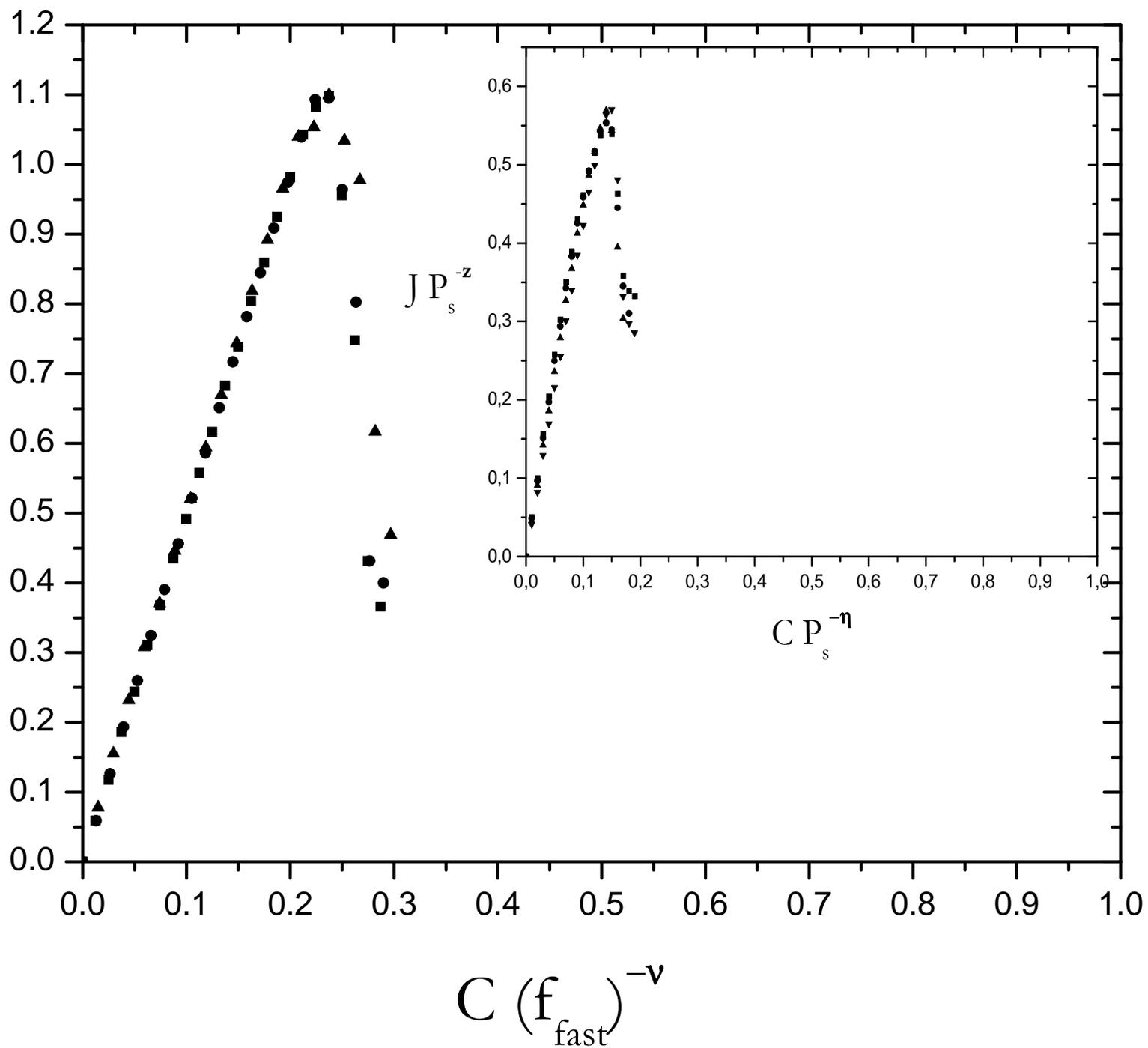
Fig.7: The scaling plot of the coexistence curve of the current J for $C\leq C_{min}$. In the main plot The simulation was done for two types of vehicles in the deterministic case $p_{s}=1$ with $Vmax{1}=5$ for type1, $Vmax{2}=1$ for type2, and for three values of $f_{fast}$: square dots correspond to $f_{fast}=0.75$, circle dots correspond to $f_{fast}=0.7$, and up triangles correspond to $f_{fast}=0.6$. In this case $\alpha=1.631\: \:and\:  \: \nu=0.773$.\\
However, in the inset The simulation was done for two types of vehicles with a fixed fraction $f_{fast}=0.75$ and for several values of $p_{s}$: square dots correspond to $p_{s}=0.3$, circle dots correspond to $p_{s}=0.4$, up triangles correspond to $p_{s}=0.5$ and down triangles correspond to $p_{s}=0.7$. In this case $z=0.302\: \:and\:  \: \eta=0$. The plot does not contain the region when $C > C_{min}$, because the system is in the absorbing phase where $f_{d}=0$ .i.e., the movement of type1 is dictated by the one of type2, which leads to the well known fundamental diagram of one species of vehicles with $V_{max}=1$.\\\
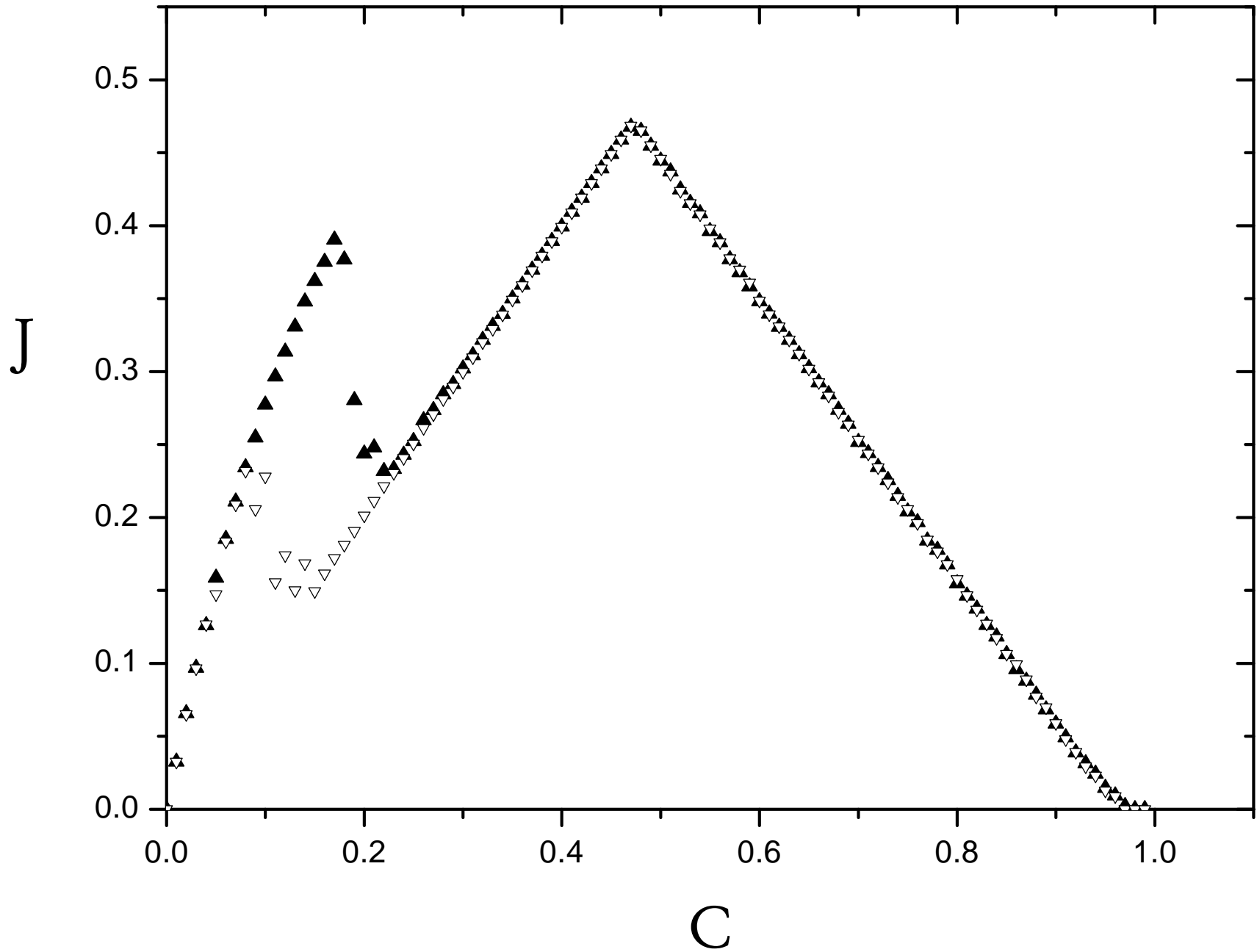
Fig.8: The fundamental diagram (J,C) for two different initial configurations: Up triangles correspond to homogeneous initial configuration and down triangles correspond to the mega jam. The simulation was done for two types of vehicles with $(Vmax{1}=5,f_{fast}=0.6)$ for type1, $(Vmax{2}=1,f_{slow}=0.4)$ for type2 and $p_{s}=0.7$.\\\
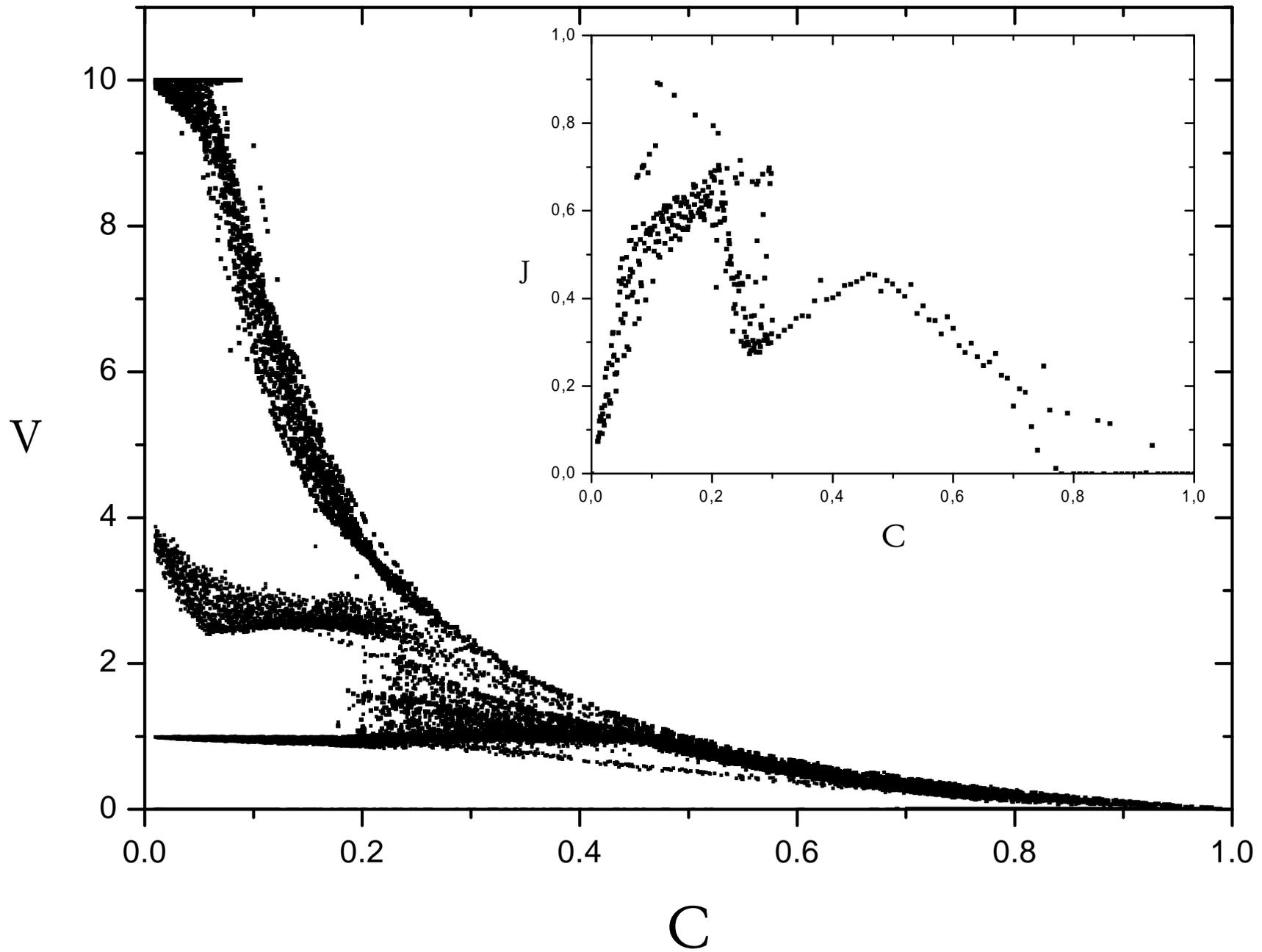
Fig.9: The fundamental diagram (J,C) and The mean speed V versus the density C (inset) for three types of vehicles in deterministic case $(p_{s}=1)$ with $Vmax{1}=10$ for type1, $Vmax{2}=4$ for type2, and $Vmax{3}=1$ for type3. For each value of the density C the fractions $(f_{fast1},f_{fast2},f_{slow})$ are chosen randomly.\\
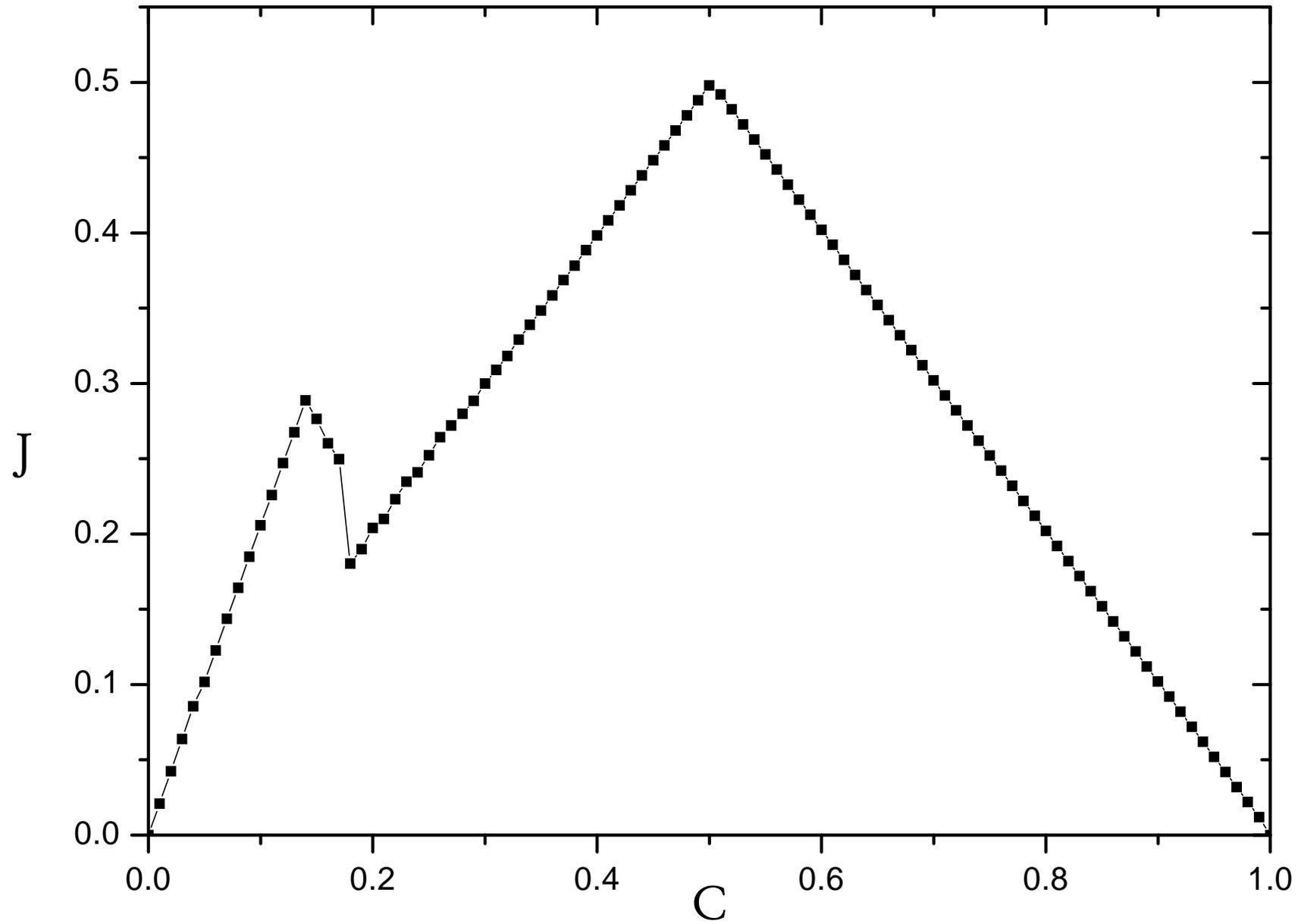
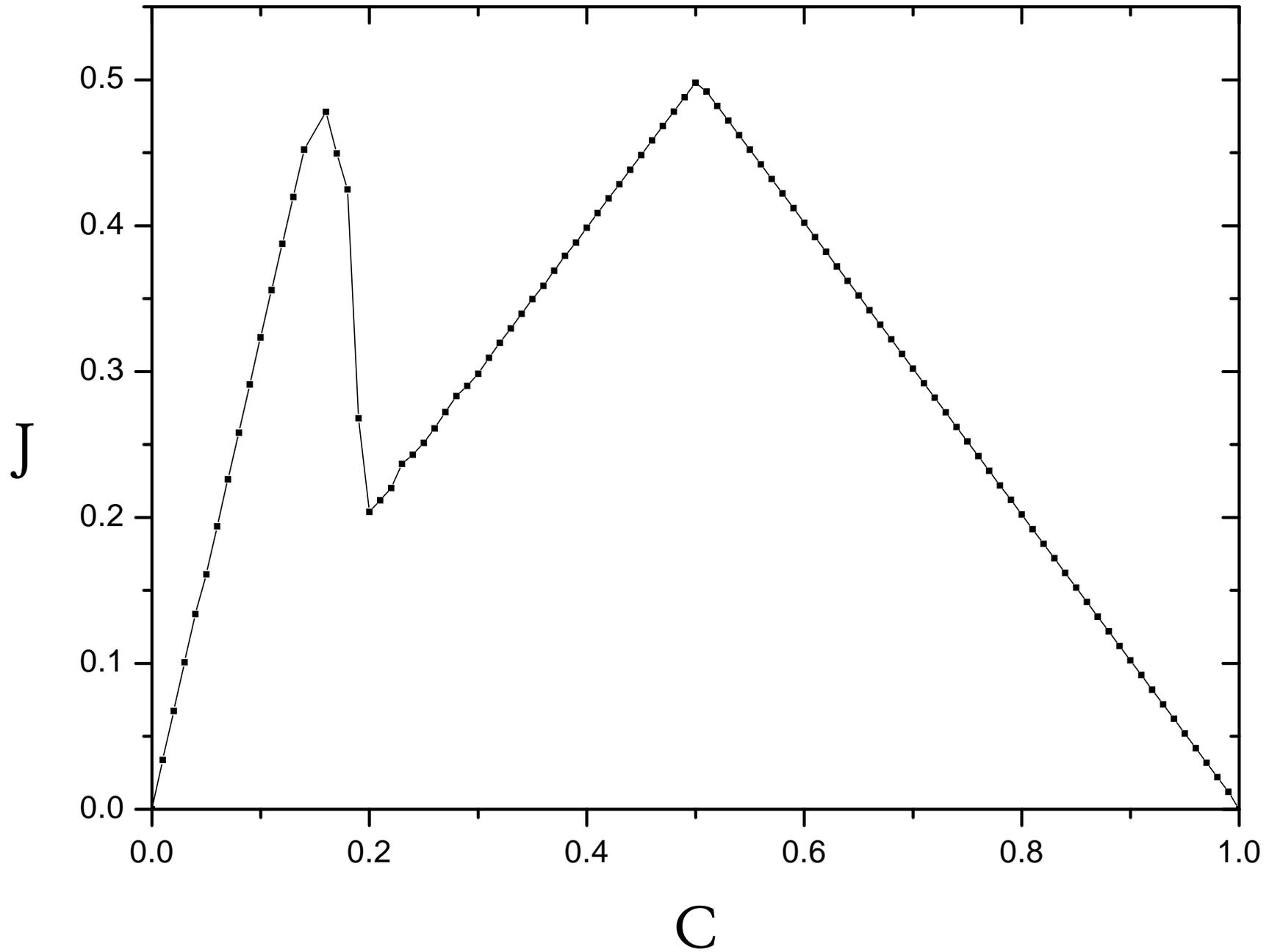
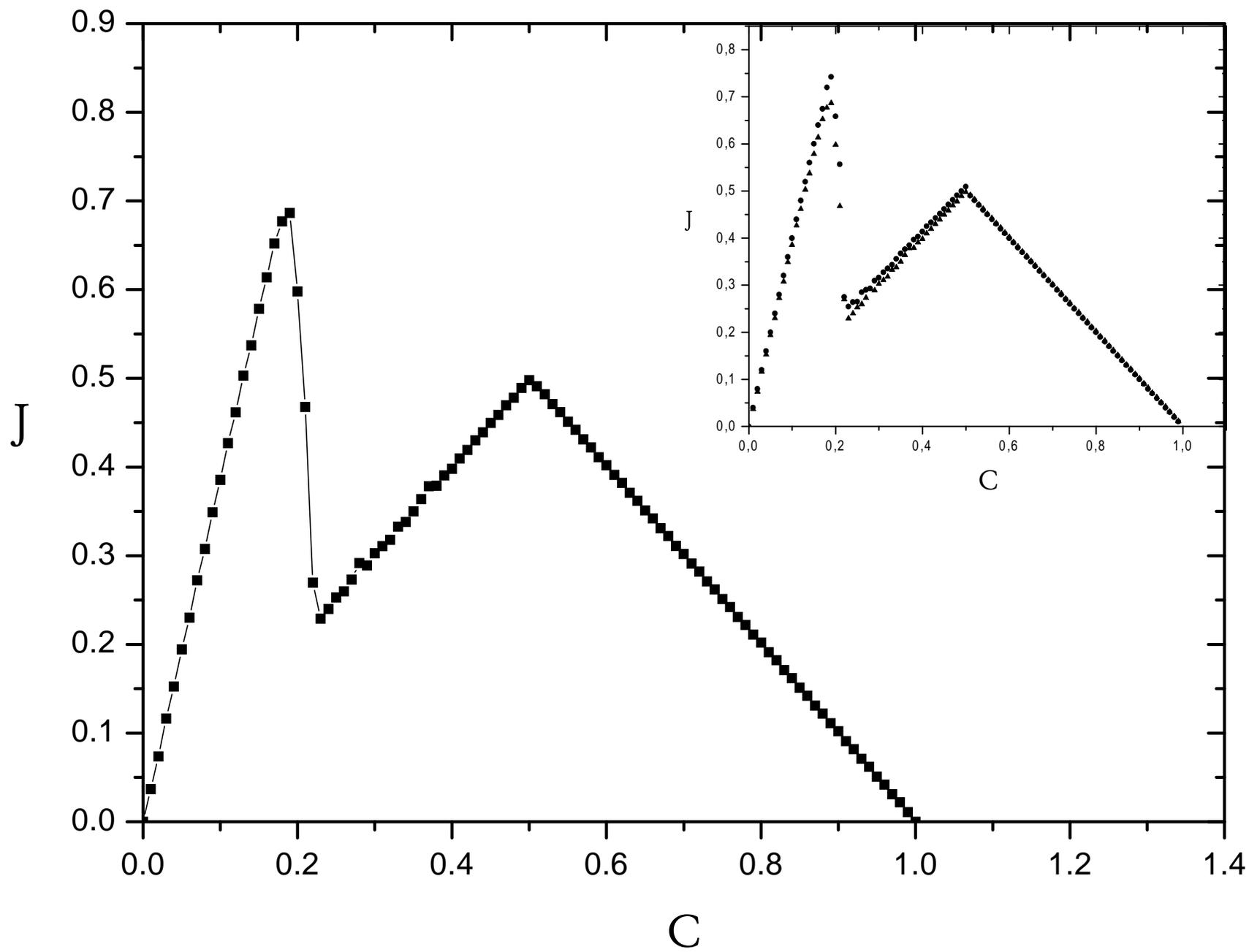
\end{document}